\documentclass{article}

\usepackage{PRIMEarxiv}

\usepackage[utf8]{inputenc} 
\usepackage[T1]{fontenc}    
\usepackage{hyperref}       
\usepackage{url}            
\usepackage{booktabs} 
\usepackage{float}
\usepackage{amsfonts}       
\usepackage{nicefrac}       
\usepackage{microtype}      
\usepackage{lipsum}
\usepackage{fancyhdr}       
\usepackage{graphicx}       
\graphicspath{{media/}}     

\title{Digital Twin Framework for Time to Failure Forecasting of Wind Turbine Gearbox: A Concept}

\author{
   Mili Wadhwani\\
  Department of Electrical Engineering \\Adani Institute of Infrastructure Engineering\\
  Ahmedabad, India\\
  \texttt{miliwadhwani.ele19@aii.ac.in} \\ \And
   Sakshi Deshmukh\\
  Department of Electrical Engineering \\Adani Institute of Infrastructure Engineering\\
  Ahmedabad, India\\
  \texttt{sakshideshmukh.ele19@aii.ac.in} \\ \And 
  Harsh S. Dhiman, Ph.D. \\
  Department of Electrical Engineering \\Adani Institute of Infrastructure Engineering\\
  Ahmedabad, India\\
  \texttt{harsh.dhiman@aii.ac.in}
  }

\begin{document}
\maketitle

\begin{abstract}\large
Wind turbine is a complex machine with its rotating and non-rotating equipment being sensitive to faults. Due to increased wear and tear, the maintenance aspect of a wind turbine is of critical importance. Unexpected failure of wind turbine components can lead to increased O\&M costs which ultimately reduces effective power capture of a wind farm. Fault detection in wind turbines is often supplemented with SCADA data available from wind farm operators in the form of time-series format with a 10-minute sample interval. Moreover, time-series analysis and data representation has become a powerful tool to get a deeper understating of the dynamic processes in complex machinery like wind turbine. Wind turbine SCADA data is usually available in form of a multivariate time-series with variables like gearbox oil temperature, gearbox bearing temperature, nacelle temperature, rotor speed and active power produced. In this preprint, we discuss the concept of a digital twin for time to failure forecasting of the wind turbine gearbox where a predictive module continuously gets updated with real-time SCADA data and generates meaningful insights for the wind farm operator.

\end{abstract}


\section*{Background}\large
The expansion of wind farms makes their operations and maintenance an important issue. According, to the data presented in, maintenance cost alone may account for a 15\%-30\% of the total generation cost. To reduce the maintenance cost, condition monitoring plays a big role, it is a cost-effective approach to analyse wind turbine performance with the help of Supervisory Control and Data Acquisition (SCADA) system which records various wind turbine parameters such as vibration analysis, temperature monitoring, etc. The main objective is to acquire information regarding the health of machines in advance through condition monitoring techniques \cite{Dhiman2021_TEC}. In such scenario digital twin can be created. A digital Twin for a wind turbine is a computer program that uses real-time data to create a virtual space of a wind farm in the physical world \cite{9696318}.

\begin{figure}[H]
    \centering
    \includegraphics[width=0.9\linewidth]{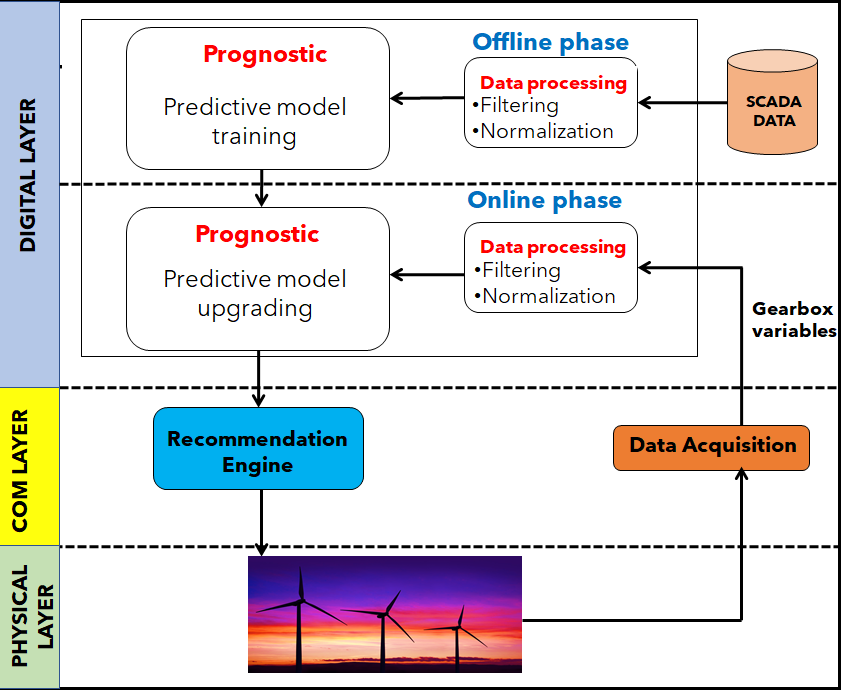}
    \caption{Schematic diagram of Digital Twin for Time to Failure Forecasting}
    \label{fig1}
\end{figure}

Concept of digital twin for a wind turbine: Digital twin can be divided into three different layers namely, physical layer, communication layer and digital layer. Physical layer embeds the physics-informed wind turbine model. Transition of decisions from digital layer to physical layer are facilitated via the communication layer. Digital layer comprises of there are two zones present namely offline and online. These layers run a predictive model that is built using historical SCADA data available from the wind turbines. Fig. \ref{fig1} illustrates such a digital twin framework. 

\section*{Methodology}
The approach for constructing a digital twin requires collection of relevant SCADA data from a high-fidelity sensor network. An instrumentation system that collects and sends data to the prognostic model as discussed above is a crucial part of the proposed digital twin. With reference to Fig. \ref{fig1}, the proposed framework would consist of a gearbox test rig equipped with precision sensors which collect real-time data for the gearbox. The data could be in form of gear oil temperature, bearing temperature, surface roughness, and high-quality images of gear tooth damages. Looking into the wide array of data being collected, the task is to implement a high-fidelity prognostic model that forecasts the remaining useful life (or time to failure) of the gearbox.

\section*{Industrial Relevance}

\begin{figure}[H]
    \centering
    \includegraphics[width=1\linewidth]{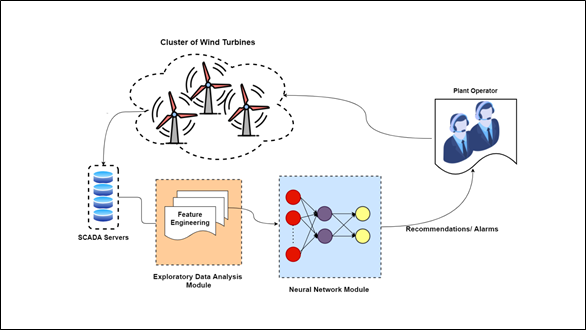}
    \caption{Industrial layout of a digital twin (including recommendation system) for wind turbine gearbox}
    \label{fig2}
\end{figure}

The market of industrial gearbox is likely to increase with a CAGR of 4.2\% over next 4-5 years and this calls for the need of an effective predictive maintenance system. In this proposal, the quality of data collected from the test bench (gearbox and bearings) is crucial and with reference to the wind turbine industry, the O\&M cost that accounts for 15-30\% of the overall cost of energy can be significantly reduced. With this approach, it is possible to develop an effective strategy to deal with turbine gearbox degradation. Time to failure forecast also involves heavy dependency on the spatio-temporal patterns across the wind turbine nacelle which can meaningful recommendations in form of alerts to the system operator as illustrated in Fig. \ref{fig2}

To understand the concept of digital twin for a wind turbine, we begin with physical layer; it is basically the layer where wind turbine is present through which we acquire data using quality sensors wherein we send data of various gearbox variables like gearbox temperature, gearbox torque etc. to the communication layer. In digital layer there are two zones namely offline and online phase. Offline phase is the phase where you create a predictive model which is a part of digital twin where you make a machine learning or deep learning model using historical (SCADA) data. Once we have SCADA data of last 10-15 years we can create a prognostic model, now to create a successful model we need to pre-process our data by filtering and normalization. Here filtering means getting rid of unwanted or say junk data from your database and using the good data we make our predictive model. Now when we deploy that predictive model with real time data, we generate some recommendations. From data acquisition, we are trying to feed our real time data to our predictive model which is running in real time in online phase. The output from this recommendation engine is to be given to the physical layer. It is a kind of decision-making zone where you give critical feedback to the physical asset. If we talk about communication layer, the digital and physical layers are connected by communication layer. The objective of the communication layer is to minimize the delay between physical and digital layer. The critical part of a wind turbine condition monitoring system is the communication network, which must be secure, robust, and reliable. Two different communication technologies, i.e., wired and wireless could be used for connection between the wind turbines and the control center. Recently, different wireless technologies have emerged, Ethernet-based, Wi-Fi-based, and ZigBee-based are among these technologies for the wind turbine communication.

\section*{Probable Outcomes}
The analysis of this research proposal is likely to deliver following outcomes and insights which are directly in harmony with the current industrial trends and could be helpful in generating critical insights for the system operator. 

\begin{itemize}
    \item Remaining useful life (RUL) estimation of the gearbox life in terms of temperature degradation.
    \item Use of advanced deep learning-based algorithms to culminate in an accurate forecast with appropriate confidence and prediction intervals.
\item RUL estimation can also generate meaningful insights about the neighbouring wind turbine components like generator, rotor blades.
\item Ready to deploy deep-learning frameworks for fault diagnosis of turbine gearboxes of varied sizes and dimensions. 
 
\end{itemize}

The Objective of digital twin is to make a virtual model which will predict the remaining useful life (RUL) of the wind turbine. It has the potential to give real-time status on performance of the machine which will increase connectivity and feedback between devices, which in turn, improves the reliability of the system. It is expected that the entire use of digital twin models in product development and production will dominate future product generations, because they allow to create high-performance products competitively

\bibliography{references}
\bibliographystyle{IEEEtran}

\end{document}